\documentclass[letter,11pt]{article}

\usepackage{xcolor}
\usepackage{float}
\usepackage{graphics,graphicx}
\usepackage{hyperref,setspace}

\begin{document}

\begin{center}
\includegraphics[width=\textwidth]{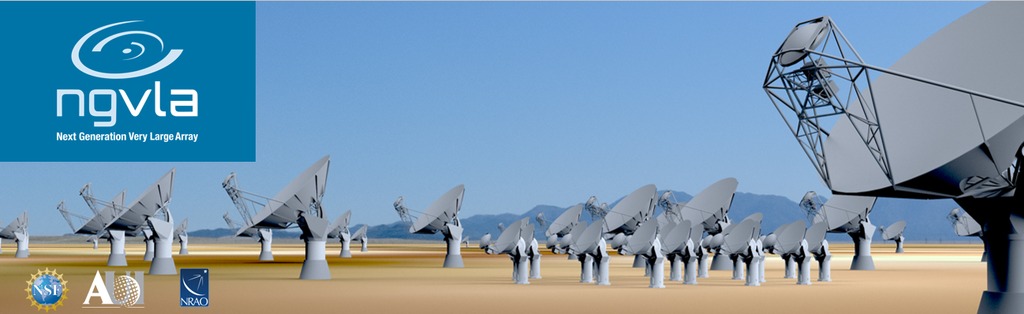}
\end{center}

\begin{center}

{\bf \large Next Generation Very Large Array Memo No. 118 \\
September 2023}

\vspace{0.1in}

{\bf \large Imaging the ring and jet in M87 at 85 GHz with the ngEHT and ngEHT+ngVLA}

\end{center}

\hrule 

\vspace{0.3cm}

\centerline{C.L. Carilli, R.C. Walker, E. Murphy, B. Mason}
\centerline{National Radio Astronomy Observatory, Socorro, NM, USA}

\vspace{0.2in}


\begin{abstract}

The Global mm-VLBI Array (GMVA) has demonstrated the ability to resolved what may be the general relativistic shadow of the supermassive black hole in M87 at 86 GHz, as well as delineate the inner jet to $\sim 1$~mas distance. We investigate the ability of the planned ngEHT, and the ngEHT + ngVLA at 85 GHz, to image such a nuclear 'ring', and the associated jet, using a constructed model based on the current estimate of the ring size, and a scaled version of the best VLBA image of the M87 jet at 43 GHz. While the resolution does not improve due to the limit set by the diameter of the Earth, the ngEHT alone should provide both a higher fidelity image of the ring on scales $\le 0.1$~mas, and a good image of a more extended jet to $\sim 1$~mas. Adding the ngVLA improves substantially the dynamic range (factor 3.5), as well as adds the ability to image structures on larger scales, in this case out to at least 5~mas, and potentially to much larger scales given the $\sim 10^5$ range in spatial scales covered by the ngVLA itself. Both arrays provide good image fidelity ($\le 0.1$), in the inner $\sim 1$~mas, but the ngEHT-only image does not reproduce the outer jet well, or at all, with fidelity values greater than unity. The combined array reproduces much of the outer jet with good fidelity ($\le 0.3$). Adding the ngVLA also decreases the susceptibility to antenna-based phase errors by a similar factor, and improves the ability for fringe fitting and subsequent phase and amplitude self-calibration. As for scales $< 100~\mu$as, ie. the ring itself, adding the ngVLA makes little change for very bright sources, where uniform weighting can be employed. But for faint sources, adding the ngVLA adds potentially an order-of magnitude sensitivity improvement (Issaoun et al. 2023).

\end{abstract}

\section{Introduction}

Global millimeter VLBI (GMVA) at 86 GHz has recently resolved the radio nucleus of the radio galaxy M87 into a ring similar to that seen at 230 GHz by the Event Horizon Telescope, but with a diameter about 50\% larger (64$\mu$as vs 42$\mu$as; Lu et al. 2023, Nature, 616, 686; EHT collaboration 2019, ApJ, 875, L1). The cause for a larger ring at lower frequency may relate to the opacity distribution of the emitting regions, although this remains speculative. 

Regardless of the reason for the larger ring at 86~GHz, the ability to resolved the radio nucleus of M87 at 86 GHz, and particular, to see what may be the result of strong field gravitational lensing by the super massive black hole of radio emission from the accretion disk, raises the possibility for the ngVLA to participate in, and potentially improve significantly, the imaging of radio AGN on scales down to tens of $\mu$as. Granted there are only two sources considered viable at this time for resolving the lensed ring (M87 and SgrA*), still, we feel it important to explore the contribution the ngVLA can make for 86 GHz global VLBI imaging on scales of tens of $\mu$as to a few milliarcseconds, or more. 

Our exploration is in the context of what may be the predominant future GMVA configuration, namely, the next generation EHT (Doeleman et al. 2023 arXiv:2306.08787). A similar ngEHT + ngVLA study has been performed by Issaoun et al. (2023 Galaxies, 11, 28), where they focus predominantly on increased sensitivity as a function of baseline length. In this memo, we consider imaging dynamic range for bright, complex sources over a wide range of scales. 

We emphasize that a full cross correlation of the ngVLA, including baselines within the Core and Spiral, as well as intra-station baselines for LONG (typically of order 100~m for a three element LONG station), has a spatial dynamic range of $\sim 10^5$, and can image sources on much larger scales (up to $\sim 10''$ at 85 GHz). This would also be true if ALMA did a full cross correlation in future ngEHT observations. Herein, we focus on scales of a few milliarcseconds down to tens of $\mu$as, as relevant for the continental and transcontinental baselines. 

\section{Model and Simulation} 

The source model is an adapted version of the VLBA high dynamic range image of M87 at 43 GHz, produced from many epochs of observation over 17 years (Walker et al. 2018, ApJ, 855, 128). 

For the jet part of the model, first a Gaussian is fit and subtracted at the core position of the 43 GHz image. The pixel size is then scaled down by a factor six to 6 $\mu$as per pixel, decreasing the overall jet size by that factor. The intrinsic resolution at 43 GHz  is about 0.3 mas, which also gets scaled by this factor down to 50$\mu$as, which can be considered a 'preconvolution' resolution of the input model of the extended jet, comparable to the resolution of the uniform weighted images obtained at 85 GHz in images below. 

The model image is then blanked by hand outside the jet, as well as blanking all pixels of negative surface brightness. Blanked pixels are then set to zero. The total flux density of the jet (minus the core) is then scale to 0.42 Jy, as per the expected jet flux density without the core at 85 GHz based on Kim et al. (2018, A\&A, 616, 18). 

The intrinsic dynamic range of the input image is $\sim 10^4$. However, the blanking process removes the dominant off-source artifacts, resulting in a model with no obvious extraneous structures. While the blanking process can be considered subjective, the fact remains that, whether using a doctored real-source model image, or a model source based on physically motivated numerical simulations, or a fabricated source of random geometric structures, the goal is simply to reproduce as closely as possible the input model, as judged by the resulting imaging artifacts and metrics. 

For the core region, a simple annulus is generated with a diameter of 64 $\mu$as and a width of 12 $\mu$as, with a total flux density of 0.68 Jy, as per Lu et al. (2023; note the annular width is not well constrained; we adopt a relatively narrow annulus). This annulus is then added to the jet model at the position of the subtracted core. 

From a single frequency image, we generate a model spectral cube of 10~GHz total bandwidth with 10 spectral channels, centered at 85~GHz. The spectral cube then allows for modest bandwidth synthesis. 

CASA SIMOBSERVE was used to generate the visibility measurement sets. A 12 hour synthesis was employed with a record length of 180 seconds.\footnote{The record length is only used for calculating the thermal noise. Otherwise, SIMOBSERVE simply adopts the measurement at the center uv-pixel of the record length, and the center of the channel width, ie. SIMOBSERVE does not simulate bandwidth or time smearing.} The data were then flagged using FLAGDATA for antenna elevations lower than 11$^o$.\footnote{Tracking the source to 0$^o$ elevation limit has a minor effect on the results, and would, in practice, add noisy data at very low elevation due to the increased pathlength through the atmosphere (a factor that is not included in the simulations).}

Two sets of antennas were employed. First was the next generation Event Horizon Telescope, as described in Doeleman et al. (2023). The adopted antennas are:\footnote{Many stations are familiar GMVA antennas; the new ones include: CNI = Canary Islands, BAJA = Baja Mexico, LAS = Coquimbo Chile, JELM = Wyoming, AMT = Namibia} 

\vskip 0.1in

\noindent {\bf ngEHT}: ALMA, APEX, SMT, OVRO, KP, GLT, JCMT, SMA, SPT, HAY, BAJA, CNI, LAS, JELM, AMT, IRAM 30m, PDBI (NOEMA), LMT

\vskip 0.1in 

For sensitivity purposes, we make a simplifying assumption by adopting an 18m ngVLA antenna in every case. For ALMA, we include 43 antennas. For larger antennas such as the LMT and the 30m, or an array such as NOEMA, we add antennas to the location to mimic the sensitivity. There are some smaller diameter antennas in the ngEHT, but these may have higher efficiency than the ngVLA antennas. In the end, the results are not limited by thermal noise  but by image reconstruction limitations, and these simulations test primarily the uv-coverage aspects of the configurations. The array has an effective collecting area of 71 18m antennas. 

We then add the ngVLA to the ngEHT configuration (RevE configurations). We include all of LONG (30 antennas in 10 stations to baselines of 8000 km), MID (46 antennas to baselines of 1000 km), and Spiral (54 antennas to baselines of 40 km). We also include 19 Core antennas to baselines of 4 km, adding an additional 149 antennas (including the full core would add an additional 95 antennas). 

Thermal noise is added to the visibilities, based on the expected sensitivity estimates in Selina et al. (2018, ASP 7, 15). Based on the total collecting area, and total bandwidth and integration time, the thermal noise for ngEHT alone would be 1.0 $\mu$Jy beam$^{-1}$ while that for the ngEHT+ngVLA is 0.33 $\mu$Jy beam$^{-1}$. However, this does not include the uv-minimum imposed in the imaging step, or the limit to the track lengths imposed by mutual visibility. Both affects will increase the noise by a factor of at most two. Regardless, the thermal noise is a factor few to ten below the measured noise in the images below, implying dynamic range limited images. We also perform simulations including $\pm 20^o$ random phase errors per antenna with a time constant = record length of 180 sec. 

Images were generated with TCLEAN using a cell size of 6$\mu$as, image size of 2048x1024 pixels, specmode = 'mfs', deconvolver = 'hogbom', niter = 20000, and gain = 0.03, with Briggs weighting with R=-2 and a uv-minimum of 30 km. A CLEAN box was set on the core and jet. The adopted image pixel size (6$\mu$as), and image size (2048 pixels), implies a uv-cell size for gridding corresponding to baselines $\le 67$~km. Hence, all baselines shorter than 67~km, which includes the ngVLA spiral and ALMA, and the intra-station baselines of LONG, are gridded into a single, central uv-cell, and hence highly down weighted when using roughly uniform weighting.  Images with and without a uv-minimum were generated, with very similar results (within a few percent), in terms of beam size, total flux density, peak surface brightness, and rms. However, the peak negative surface brightness increased by a factor two without a uv-minimum, and more substantial large scale artifacts could be seen in the images, suggesting that, even with R=-2, the many many short baselines on tens of meters to 1 km scales are not benefiting the imaging for a source that is only 5~mas in extent. A multiscale CLEAN was also employed with little difference. 

\subsection{Results}

\subsubsection{UV-coverage and synthesized beams}

Figure 1 shows the uv-coverage of the ngEHT and the ngEHT+ngVLA. For the very longest spacings, beyond 8000 km (2 G$\lambda$), the uv-coverage is not dramatically different. This is because, in the E-W direction the longest baselines are short tracks of mutual visibility between Hawaii and Europe, while in the N-S direction the longest baselines are between Greenland and Chile. There is a very long baseline between the Southwest USA and Namibia (11,000 km). These baselines show substantially improved coverage adding the ngVLA. These spacings will also have dramatically improved sensitivity, by a factor of roughly 25, based on the relative number of antennas involved. For baselines from zero out to 4000 km N-S, and 8000 km E-W, adding the ngVLA makes a dramatic difference in the uv-coverage, leading to almost uniform uv-coverage over this wide range, as opposed to the substantial holes and gaps for the ngEHT only. Again, by including the ngVLA, the raw sensitivity is improved by a factor of at least three, depending on the relative contribution of ALMA to a given uv-region.\footnote{For a detailed analysis of the changes in sensitivity and the number of antennas on a given baseline for the ngEHT with and without the ngVLA, see:  presentations by Roelofs et al. in 'Broadening Horizons', and Issaoun et al. 2023} 

The resulting synthesized beams (point spread function) are shown in Figure 2. The Gaussian fit FWHM values are: $52\times 45$ $\mu$as at $67^o$ for the ngEHT only, and $54\times 46$ $\mu$as at $50^o$ for the ngEHT+ngVLA. The ngEHT-only PSF has a peak negative sidelobe of -0.19 and positive sidelobe of +0.18. The ngEHT+ngVLA has peak sidelobes of -0.15 and +0.19. Hence, the peak sidelobes are not dramatically different, but the ngEHT-only PSF shows higher sidelobes (by about 30\% to 50\% relative to the ngEHT+ngVLA PSF, on average) extending about a factor two further from the peak (to 0.3mas), as well as a more substantial inner negative ring corresponding to the first negative sidelobe, by a similar factor. These differences can be traced to the gaps in the uv-coverage on various scales seen in Figure 1. 

\subsubsection{Dynamic Range}

The resulting images of the full jet+core at $\sim 50$ $\mu$as resolution are shown in Figure 3, including the input model convolved with the ngEHT PSF. These images include thermal noise. We discuss phase errors below. 

The total flux density, peak surface brightness, peak negative, and off-source rms noise in the images are shown in Table 1. The dynamic range (peak/rms) of the ngEHT-only image is 10,400, while that of the ngEHT+ngVLA image is 37,000.

\begin{table}
\centering
\footnotesize
\caption{Imaging results}
\begin{tabular}{lcccc}
  \hline
  \hline
Array & Total & Peak & Minimum & rms \\
~  & Jy & Jy beam$^{-1}$ &  mJy beam$^{-1}$ & $\mu$Jy beam$^{-1}$ \\
  \hline
ngEHT & 1.046 & 0.211 & -0.13 & 20 \\ 
ngEHT + ngVLA & 1.106 & 0.222 & -0.12 & 6 \\
ngEHT $\delta\phi = 20^o$ & 0.970 & 0.203 & -0.86 & 220 \\
ngEHT + ngVLA $\delta\phi = 20^o$ & 1.106 & 0.217 & -0.40 & 100 \\
\hline
\hline
\vspace{0.1cm}
\end{tabular}
\label{table:possible_sites}
\end{table}

The factor three or so higher noise in the ngEHT-only image can be traced to the large scale artifacts parallel to the jet direction. The ngEHT-only image recovers poorly structure on scales $\ge 0.5$ mas, and the outer jet is essentially lost. The ngEHT+ngVLA recovers these structures reasonably well (see Sec. 2.1.3). This difference can be traced to the much better uv-coverage on baselines $< 4000$ km for the ngEHT+ngVLA. 

Figure 4 shows images of the core region, namely the ring, in hyper-resolved images, meaning TCLEAN images restored with a 37~$\mu$as Gaussian beam, as per Lu et al. (2023). A ring, meaning a central depression, can be discerned although, as in Lu et al., the width may be difficult to determine at this resolution, and there are false peaks and minima around the ring at the $\pm 15\%$ level, indicating the limits to which sub-structure could be determined. 

There is little difference between the ngEHT-only and the ngEHT+ngVLA image on these very small scales. This similarity is due to the fact that information on these scales ($< 100\mu$as or baselines $> 8,000$~km), comes from the outer-most baselines, for which the uv-coverage is not so different between the ngEHT-only and the ngEHT-ngVLA (Fig 1). In general, for bright sources, where sensitivity is not a factor and uniform weighting is employed, we expect similar images. For faint sources, the ngVLA to Namibia baselines ($\sim 11,000$~km), do add a major boost in sensitivity on these very small scales, as discussed in Section 2. 

\subsubsection{Fidelity} 

The image fidelity is shown in Figure 4, defined as: (Model - Image)/Model, where the Model has been convolved with the Gaussian CLEAN beam. Hence, values close to zero are good, and the values are fractional errors. The model is blanked at 3$\sigma$ of the cleaned image noise for the ngEHT+ngVLA image. 

For the ngEHT only, the inner $\sim 1$~mas of the jet is recovered well, with fidelity values $\le 0.1$ in absolute value. However, beyond this distance, the outer jet is poorly recovered, or not at all, with absolute values of fidelity $> 1$ in most areas, except along the brighter edges. 

For the ngEHT+ngVLA, much of the outer jet to 5~mas distance is recovered reasonably, with absolute value of fidelity $\le 0.3$. The values are systematically below zero, which may be a consequence of the weighting scheme and subsequent gridding used by TCLEAN. We tested if this was due to the uv-minimum and obtained a similar result. We will explore imaging larger sources with the full range of baselines and different methods for recovering large scale structure in future work. 

\subsubsection{Phase errors and Calibration}

We added random antenna-based phase errors with a flat distribution over $\delta\phi = \pm 20^o = \pm 0.35$rad to the visibilities. This error distribution corresponds roughly to a Gaussian with a FWHM $\sim 20^o$. The time constant is per record, or 180 sec, implying M = 240 time records. The expected image dynamic range for random antenna-based phase errors (Perley 1999, SIRA II p. 275) for an array of N elements is:

$$ D = M^{1/2}N/\delta\phi_{rad}$$

For ngEHT-only (N=71), the dynamic range limit due to phase errors is then 3100, while for the ngEHT+ngVLA (220 antennas), the value is 9700.  These can be considered an upper limit to the dynamic range in the case considered herein, since imposition of a uv minimum in TCELAN, as well as the finite uv-cell size, implies that short baselines are either not used, or highly down-weighted, in the imaging (see section 2).  

The resulting total flux density, peak surface brightness, peak negative, and off-source rms noise in the ngEHT-only image are listed in Table 1.  The dynamic range of the ngEHT-only image with phase errors is 922, while that of the ngEHT+ngVLA image is 2200.

Hence, the increased number of antennas does improve the susceptibility to phase errors by about the expected factor dictated by the relative N, but the results should be considered qualitative, for the reasons given above, and since the adopted arrays are 'unnatural' in that individual antennas larger than 18m were approximated by multiple 18m antennas. Further, the nature of the phase errors (eg. random vs. constant in time and frequency), changes the relationship between dynamic range and phase errors (see Perley 1999). 

A more general point concerning calibration is that more antennas and greater sensitivity improves the ability to calibrate, both in terms of the initial fringe fitting to remove antenna-based delay offsets in order to then average in time and frequency, while also improving the ability for subsequent self-calibration to remove residual antenna-based phase and amplitude errors due to the weather or clock errors or electronic gain instabilities. Two capabilities in particular may be of interest: 

\begin{itemize}
    \item Many of the ngVLA elements are actually 'stations', with 3 or more antennas. Stations allows for the possibility of paired antenna phase calibration (eg. Zauderer et al. 2016, AJ, 151, 18). 
    \item The addition of many intermediate and shorter baselines (a few hundred km), pairing to almost all the antennas in the array provides critical cross-checks on the amplitude scale for visibilities on the longer baselines, where amplitude self-calibration may often be only marginally constrained (Walker 2020, ngVLA memo 84).
\end{itemize}

\section{Conclusions}

The future ngEHT at 86 GHz should have the ability to resolve the ring in M87 (as per current GMVA observations), as well as image the inner jet to $\sim 1$~mas distance. Adding the ngVLA improves substantially the dynamic range (factor 3.5), as well as adds the ability to image structures on much larger scales, in this case out to almost 5~mas, but potentially much larger. Both arrays provide good image fidelity in the inner $\sim 1$~mas, but for the outer jet, the ngEHT-only image provides poor, or no, fidelity ($\ge 1$ in absolute value), while the combined array provides reasonable fidelity ($\le 0.3$ in absolute value). 

Adding the ngVLA also decreases the susceptibility to antenna-based phase errors by a similar factor, and improves the ability for fringe fitting and subsequent phase and amplitude self-calibration. 

As for scales $< 100~\mu$as, ie. the ring itself, adding the ngVLA makes little change for very bright sources, when uniform weighting can be employed. But for faint sources, adding the ngVLA adds potentially an order-of magnitude sensitivity improvement (Issaoun et al. 2023).

In a future memo, we will explore the contribution of the ngVLA to space VLBI at 85 GHz. Space VLBI would increase the resolution substantially (factor of 3, or potentially much more; Fromm et al. A\&A, 2021, 649, 116; Lazio et al. 2020 arXiv:2005.12767, Gurvits 2000, Advances in Space Research, 26, 739;  Kim et al. 2023, ApJ, 952, 34), and hence increase the number of target radio AGN for which the strong lensing structure may be resolved (Fish et al. 2020, Advances in Space Research, 65, 821; 	Haworth et al. 2019, arXiv:1909.01405). The expectation is that the vastly increased collecting area dispersed on continental scales should make substantial improvements to Space VLBI sensitivity, imaging, and calibration. 

\clearpage
\newpage

\begin{figure}
\centering 
\centerline{\includegraphics[scale=0.3]{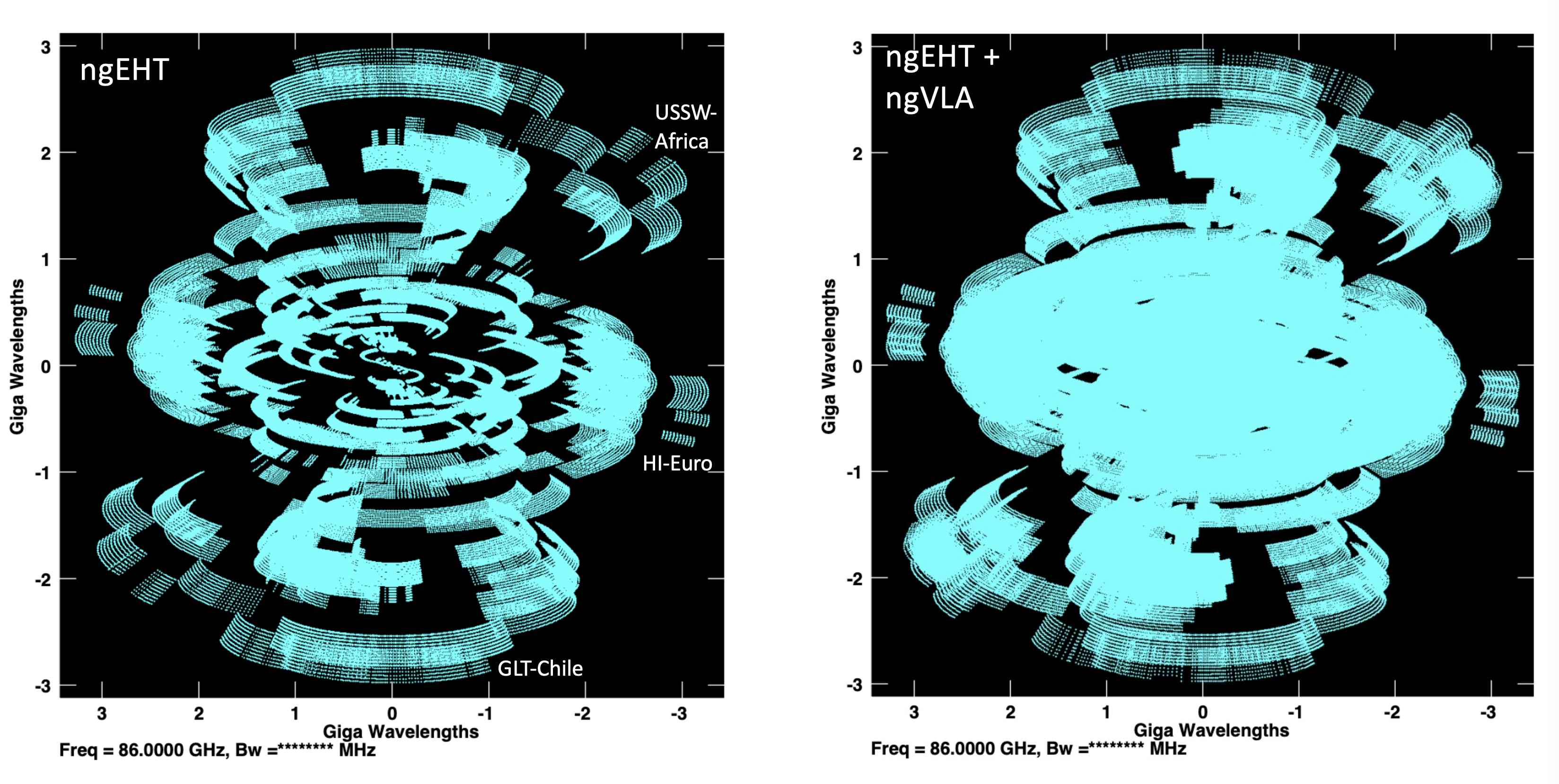}}
\caption{The UV coverage for the 12 hour synthesis including bandwidth synthesis from 80 GHz to 90 GHz, for the adopted configuration. A few characteristic antenna pairs for the longest baselines ($>$10,000 km) are shown.
}
\end{figure}  

\begin{figure}
\centering 
\centerline{\includegraphics[scale=0.4]{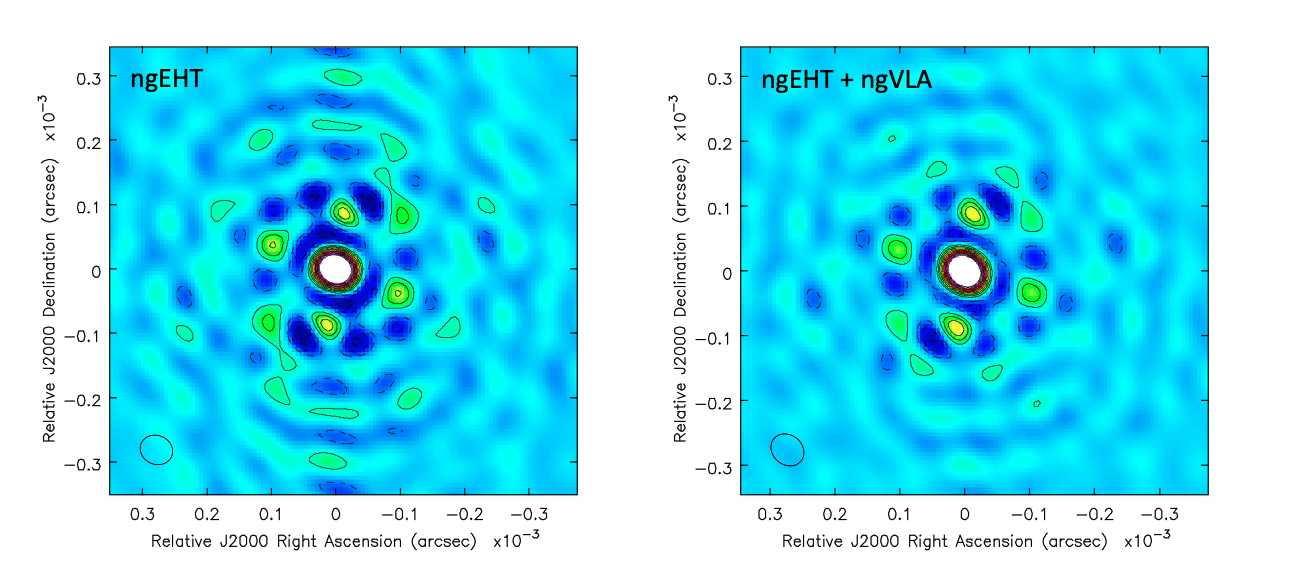}}
\caption{Left: the PSF for the ngEHT for a 12 hour synthesis at 85 GHz with a 10\% bandwidth. The contour levels are linear starting at 0.05 in increments of 0.05. The peak contour is the 50\% point of the synthesized beam. Negative contours are dashed. Right: same but for ngEHT + ngVLA. 
}
\end{figure}  

\begin{figure}
\centering 
\vspace*{-4cm}
\centerline{\includegraphics[scale=0.32,angle=-90]{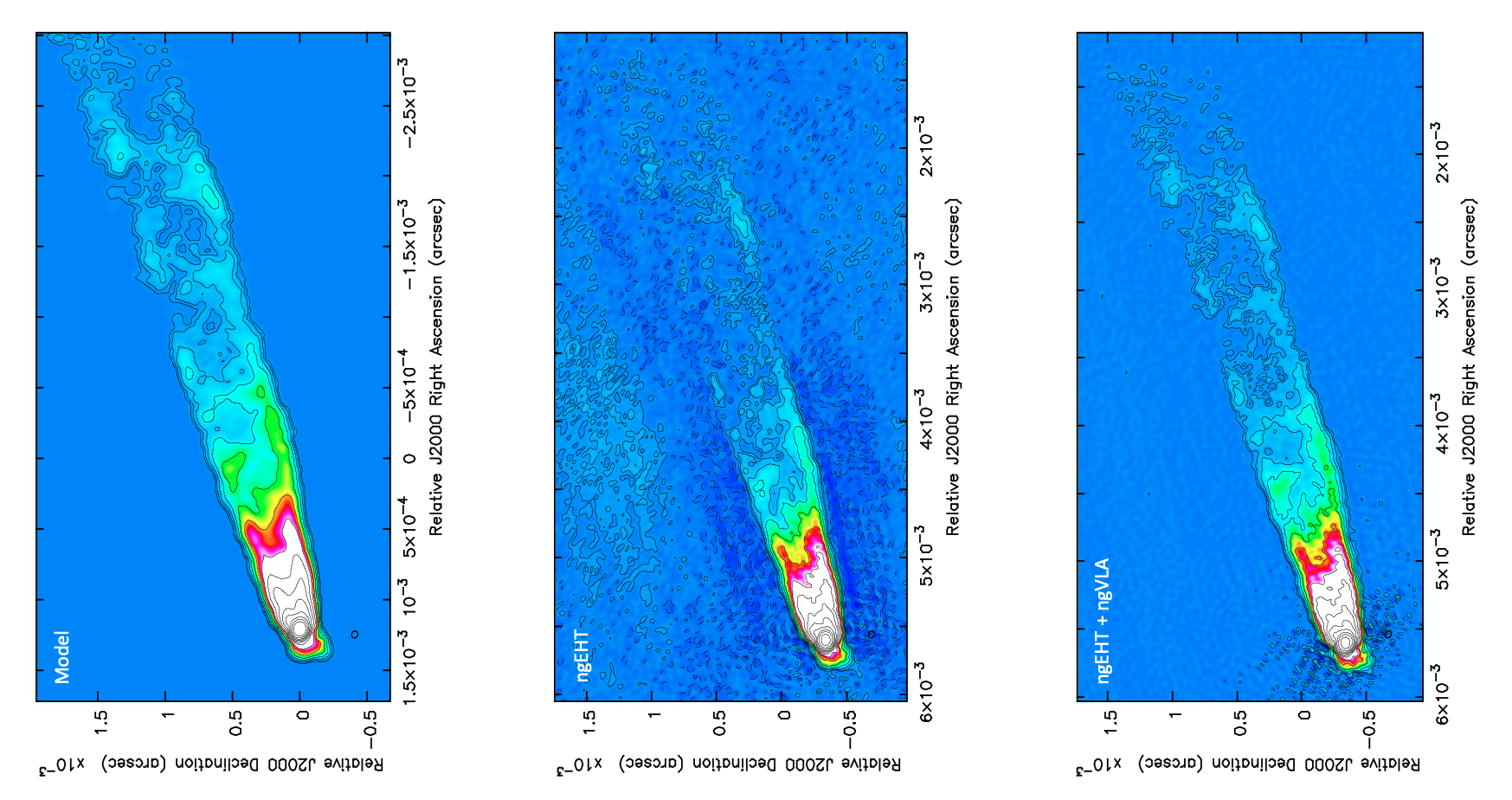}}
\caption{Top: Input model of M87 at 85 GHz built from the VLBA 43 GHz image as described in Section 2, and convolved with the ngEHT Gaussian beam. Middle: ngEHT only image for a 12 hour synthesis. The synthesized beam size is: $52\times 45$ $\mu$as at $67^o$. Bottom: ngVLA + ngEHT image for a 12 hour synthesis. The synthesized beam size is: $54\times 46$ $\mu$as at $50^o$. The contour levels in both cases are a geometric progression in factor two starting at 20 $\mu$Jy beam$^{-1}$. The color scale is the same in both images. 
}
\end{figure}  

\begin{figure}
\centering 
\centerline{\includegraphics[scale=0.3]{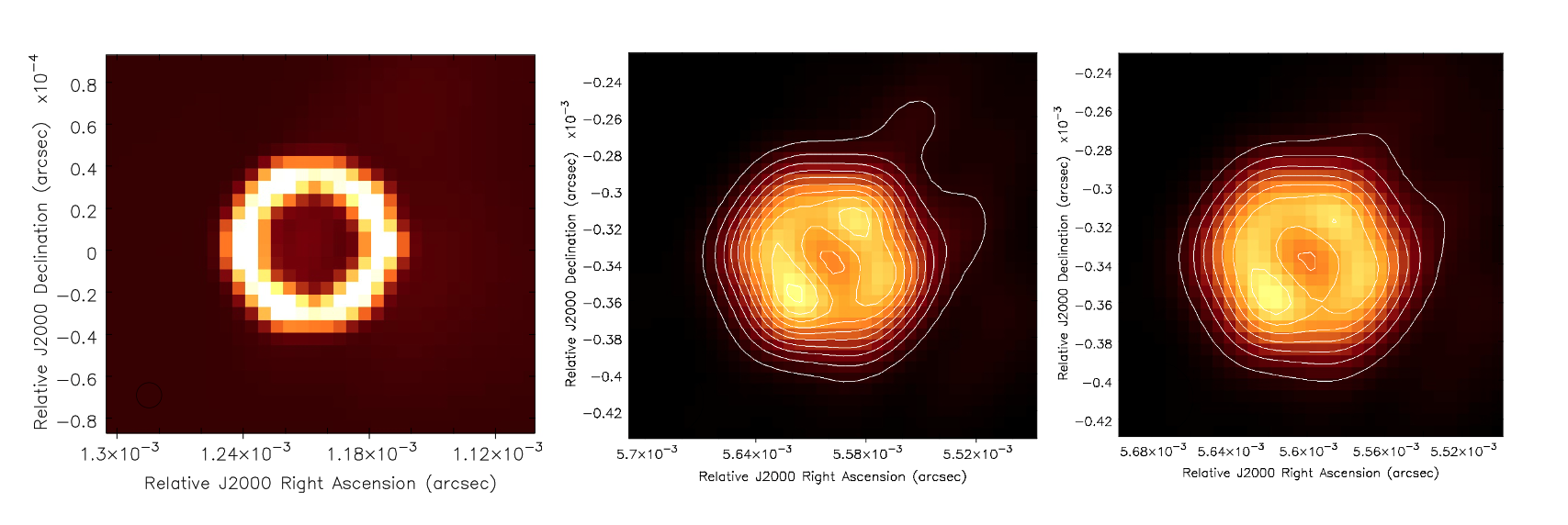}}
\caption{Left: Input model of M87 core region at 85 GHz based on the ring seen in the GMVA image of Lu et al. (2023). Middle: ngEHT image hyper-resolved by restoring with a Gaussian beam of FWHM 37$\mu$as, as per Lu et al. (2023). Right: hyper-resolved ngVLA + ngEHT image.
The contour levels are linear starting at 15 mJy beam$^{-1}$ in increments of 15 mJy beam$^{-1}$. The color scale is the same in both images.
}
\end{figure}  

\begin{figure}
\centering 
\vspace*{-2cm}
\centerline{\includegraphics[scale=0.4,angle=-90]{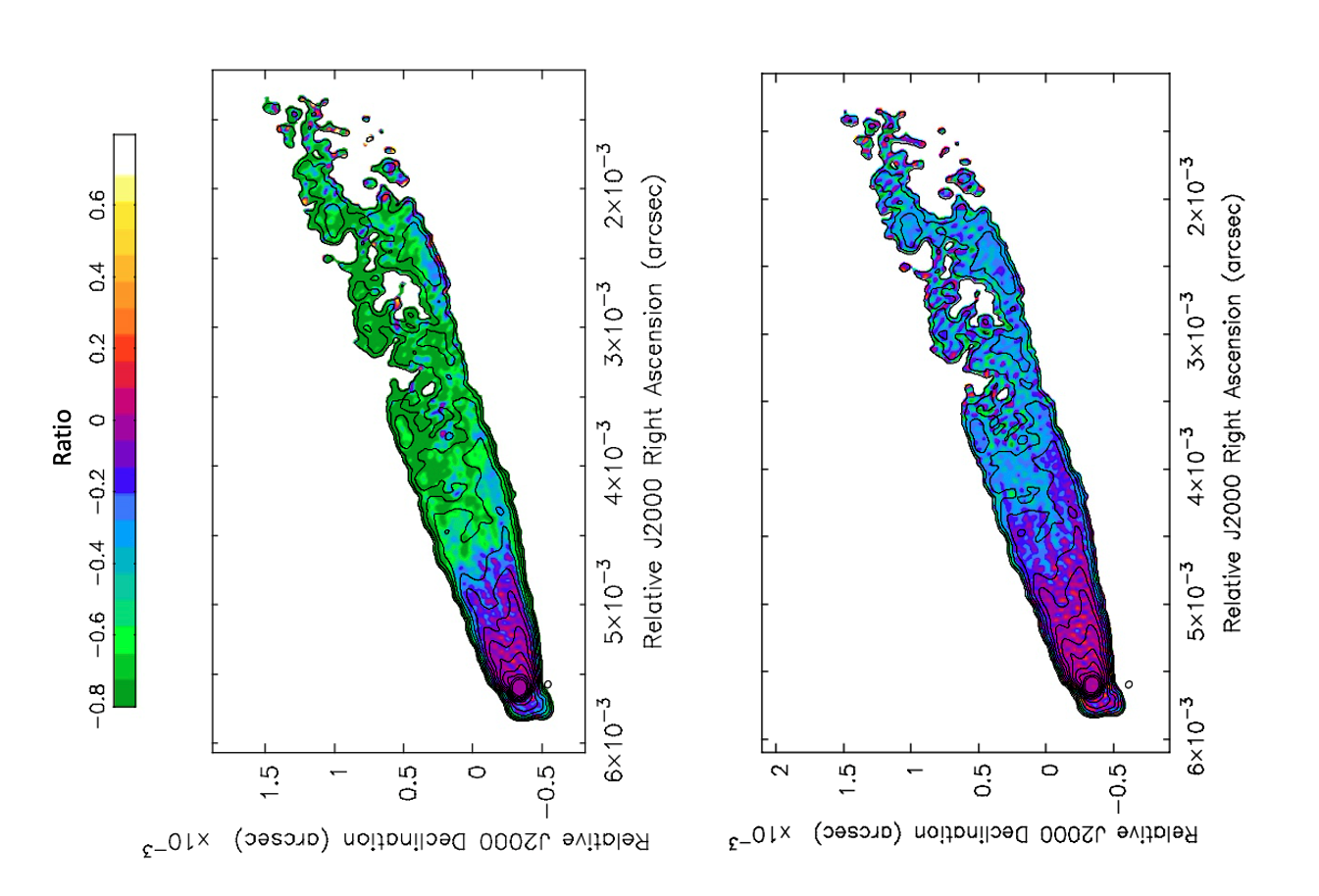}}
\caption{Fidelity images as defined in section 2.1.3. Top: ngEHT-only. Bottom: ngEHT+ngVLA. The contour levels are in factors of root two of the smoothed model image starting at 20$\mu$Jy beam$^{-1}$. 
}
\end{figure}  

\end{document}